# Achieving Differential Privacy against Non-Intrusive Load Monitoring in Smart Grid: a Fog Computing approach

Hui Cao[1], Shubo Liu[1*], Longfei Wu[2], Zhitao Guan[3], Xiaojiang Du[4]

[1](*School of Computer, Wuhan University, Wuhan 430072 China*)

[2](*Department of Mathematics and Computer Science, Fayetteville State University, Fayetteville, NC 28301 USA*)

[3](*School of Control and Computer Engineering, North China Electric Power University, Beijing 102206 China*)

[4](*Department of Computer and Information Sciences, Temple University, Philadelphia, PA 19122 USA*)

*Abstract*—Fog computing, a non-trivial extension of cloud computing to the edge of the network, has great advantage in providing services with a lower latency. In smart grid, the application of fog computing can greatly facilitate the collection of consumer's fine-grained energy consumption data, which can then be used to draw the load curve and develop a plan or model for power generation. However, such data may also reveal customer's daily activities. Non-intrusive load monitoring (NILM) can monitor an electrical circuit that powers a number of appliances switching on and off independently. If an adversary analyzes the meter readings together with the data measured by an NILM device, the customer's privacy will be disclosed. In this paper, we propose an effective privacy-preserving scheme for electric load monitoring, which can guarantee differential privacy of data disclosure in smart grid. In the proposed scheme, an energy consumption behavior model based on Factorial Hidden Markov Model (FHMM) is established. In addition, noise is added to the behavior parameter, which is different from the traditional methods that usually add noise to the energy consumption data. The analysis shows that the proposed scheme can get a better trade-off between utility and privacy compared with other popular methods.

## I. INTRODUCTION

With the support of emerging information technologies like the Internet of Things (IoT), fog computing, and cloud computing, smart grid has become increasingly intelligent and efficient [1-3]. As shown in Figure.1, in smart grid, IoT devices have been widely used for data collection purposes, such as the electricity consumption information gathering, the transmission line monitoring, and the transformer substation monitoring [4, 5]. The data collected by IoT devices are aggregated by the fog nodes that in the form of gateways or data aggregators with on-board computing capabilities; then the aggregated data are transferred to the control center for further analysis. For instance, in the electricity consumption information gathering scenario, to optimize the energy utilization, lots of smart meters (SM) installed at users' households are connected to the communication network. They can send their power consumption data to the control center via the fog layer comprised of multiple aggregators at a fine granularity [6].

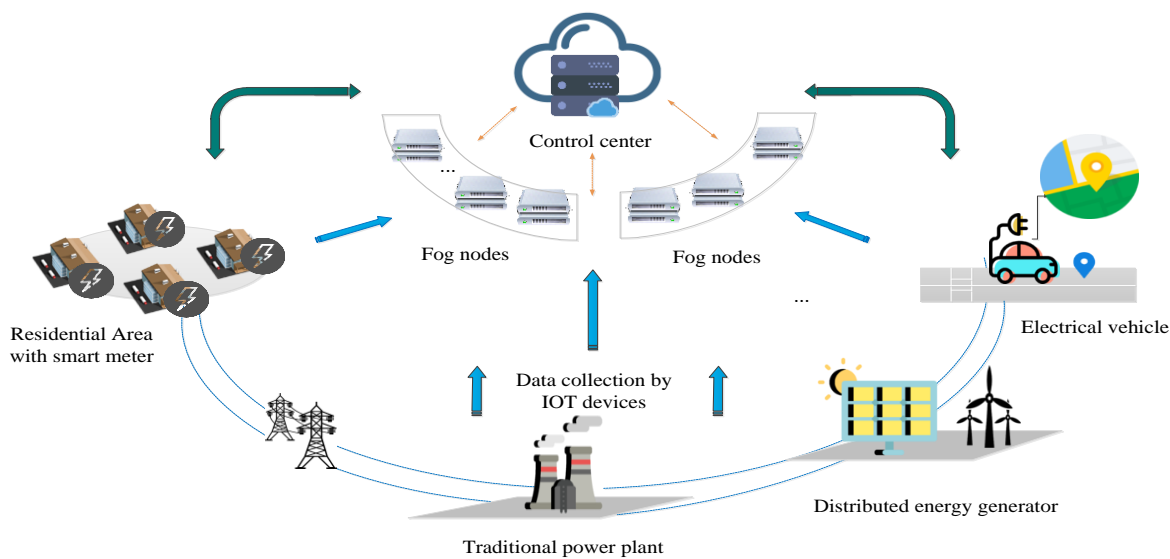

Figure.1 Fog computing enabled data collection in smart grid

However, accurately collecting user data always induces privacy and security issues as studied in [14,15,32,39]. The fine-grained energy consumption data collected by smart meters may disclose the sensitive information regarding the power consumption patterns of the household appliances which raises serious concerns about the user's privacy [7]. As shown in Figure.2, non-intrusive appliance load monitoring (NILM) is an advanced power signature analysis tool, which is often used to break down the aggregate energy consumption data into individual appliances [8]. Given a user's load profile, an adversary can track the states (ON or OFF) of all appliances with NILM. Based on the extracted device-level energy consumption data, the adversary can further infer lots of privacy-sensitive information about the user's habits and behaviors. For example, the adversary can figure out whether there is nobody at home in a specific period of time, when the users go to bed and get up, when the users leave for work and so on. Therefore, users require a privacy-friendly scheme to protect such privacy-sensitive information.

In fact, there is a significant body of work analyzing the users' privacy-preservation [9-13]. The major privacy-preserving solutions can be classified into homomorphic encryption [42] [43], flattening energy signatures by battery-based load hiding (BLH) [16-21] and noise addition [22-24]. However, schemes based on homomorphic encryption have huge computational cost and require a third party for key distribution and management [24]. These solutions are infeasible when used in a wider area with a large number of meters. However, the credibility of the third party is difficult to guarantee. The privacy-preserving schemes based on rechargeable battery are limited to the battery capacity. Moreover, the charging and discharging of the household battery may conflict with the user's economic interest [25]. Installing batteries and equipment in each home is infeasible. Zhao [1] adopts the BLH method to preserve user's privacy-sensitive information and uses differential privacy to measure the privacy-preserving performance. Noise addition is a common solution to provide differential privacy in which the outcome is not significantly affected by the removal or addition of single participants.

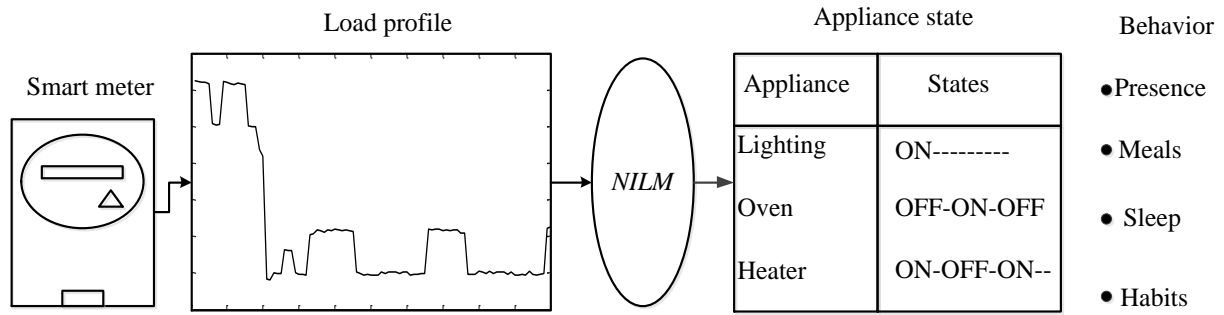

Figure.2 Non-intrusive appliance load monitor and user's behavior privacy inference in smart grid

Differential privacy is a better solution to the problem that existed methods such as k-anonymity are not sufficient to guarantee the anonymity of users. It requires that adding noise into the statistical results according to the sensitivity of each statistic when publishing the statistical results of the dataset. Therefore, whether or not the individual is in the dataset, the statistical result will not be affected. Differential privacy has been widely used in the fields of data statistics, information publishing, information searching and data mining, such as recommendation system, trace analysis and so on.

However, most schemes based on differential privacy are mainly used to protect individual information for a statistical dataset, and existing schemes applying differential privacy to smart grid have several problems. For example, Barbosa [24] proposed a fundamental work on the application of differential privacy in smart grid. Nevertheless, the trade-off between utility and privacy is not very ideal in his scheme. Therefore, designing a reasonable data obfuscation algorithm by noise addition with a better trade-off between utility and privacy is the focus of our paper. We summarize our contributions as follows:

1) Differing from traditional differential privacy schemes, we add noise into the switch states of each appliance to provide differential privacy.

2) We use the basic properties of differential privacy to prove the effectiveness of our scheme in privacy-preservation.

3) Motivated by the lower bound on utility which is called discriminant proposed by Kifer [25], we define a measurement to prove the better performance of our scheme in data-utility.

4) We adopt the information theory of differential privacy proposed by Cuff [26] to measure the trade-off between utility and privacy in our scheme.

The rest of this paper is organized as follows. Section II introduces the background and related work. In section III, our scheme is stated. In section IV, security analysis is given. In Section V, the performance of our scheme is evaluated. In Section VI, the paper is concluded.

II. BACKGROUND AND RELATED WORK

*A. Differential privacy*

Dwork [27] has proposed the notion of differential privacy for general datasets and presented how to realize differential privacy by adding noise [28]. Using the infinite divisibility of Laplace Distribution to provide differential privacy in smart grid was discussed in [29] [30]. McSherry [31] studied the parallel composition and stable transformation in differential privacy. Kifer [25] analyzed the privacy-utility tradeoff and provided the metrics for data-utility. For the differential privacy in smart grid, Won J [44] analyzed the fault-tolerance during the data aggregation and used differential privacy to protect the future ciphertext. Shi [33] applied differential privacy to preserve the metering data during the data-aggregation. Several other papers (e.g., [34-37]) have studied related security and network issues.

1) Definition of differential privacy

$M$ is a randomized algorithm. For any datasets $D_i$ and $D'$ differing from at most one element, and all subsets of possible answers $S \subseteq Range(M)$, M satisfies $\varepsilon$-differential privacy if both of the datasets satisfy the following condition:

$$P_r\{M(D_i) \in S\} \leq e^\varepsilon \times P_r\{M(D') \in S\} \quad (1)$$

The smaller the value of $\varepsilon$ is, the higher the degree of privacy-preservation is.

2) Property1: Parallel Composition

$M_1, M_2 ... M_n$ are different randomized algorithms with the privacy budgeting parameters $\varepsilon_1, \varepsilon_2 ... \varepsilon_n$. Then, the combined algorithm $M(M_1(D_1), M_2(D_2) ... M_n(D_n))$ provides $(\max \varepsilon_i)$-differential privacy for the disjoint datasets $D_1, D_2 ... D_n$.

3) Property2: Stable Transformations

For any two databases $E$ and $F$, we say T provides c-stable if it meets the following condition.

$$|T(E) \oplus T(F)| \leq c \times |E \oplus F| \quad (2)$$

$\oplus$ represents the XOR operation. If the privacy preserving mechanism M provides $\varepsilon$-differential privacy and $T$ is a c-stable transformation, the combination $M$ and $T$ provides $(\varepsilon \times c)$-differential privacy.

*B. Hidden Markov Model*

In order to get a better trade-off between the privacy-preservation and data-utility, Sanker [23] adopts the Hidden Markov Model (HMM) to model the state sequences of the appliances and generates the new energy consumption data based on the estimated state sequences and HMM. Besides, a noise following normal distribution is added to the new energy consumption data for further obfuscation. Based on the Factorial Hidden Markov Model (FHMM) [36], Kim [37] proposes a Conditional FHMM (CFHMM) to estimate the hidden states of each appliance.

1) Hidden Markov Model

Hidden Markov Model is a finite model that describes a probability distribution over sequential data. As shown in Figure.3 (a), X denotes the hidden states during different times, which can be viewed as a Markov Model and satisfies $P(x_{t+1} | x_t, x_{t-1},...x_1) = P(x_{t+1} | x_t)$. Y denotes the observed states decided by the hidden states. There are three important parameters in HMM model: initial state probability distribution, transition matrix and emission matrix.

The initial states probability distribution can be described as follows $\pi = \{\pi_i | \pi_i = P(x^{(i)}_1), 1 < i < N\}$. The transition matrix can be described as follows $A = \{P(x^{(i)}_{t+1} | x^{(i)}_t), 1 < i < N\}$. Given a discrete or a continuous set, the emission matrix can be described as $B = \{P(y^{(i)}_t | x^{(i)}_t), 1 \leq i \leq N\}$, representing the probability of emission of observed state $y_i$ when the hidden state is $x_i$.

2) Factorial Hidden Markov Model

Hidden Markov Model is a finite model that describes a probability distribution over sequential data. X denotes the hidden states during different times, which can be viewed as a Markov Model and satisfies $P(q_{t+1} | q_t, q_{t-1},...q_1) = P(q_{t+1} | q_t)$. Y denotes the observed states decided by the hidden states. There are three important parameters in HMM model: initial state probability distribution, transition matrix and emission matrix.

As an extension of HMMs, FHMM is used to model multiple independent hidden state sequences in different times. The structure is shown in Figure.3 (b). $X_i$ represents independent hidden states sequence. $Y_i$ represents corresponding observed states sequence. $Y_{sum}$ represents the aggregated observed states sequence.

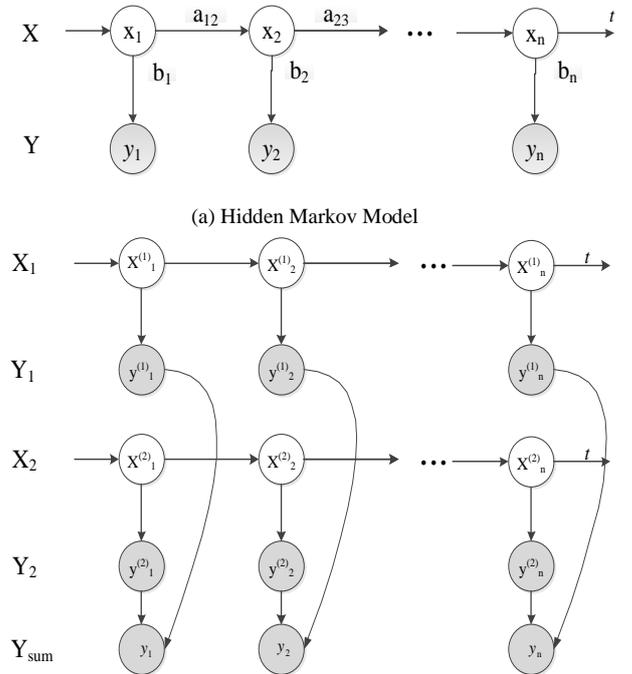

(a) Hidden Markov Model

(b) Factorial Hidden Markov Model
Figure.3 Sample graph of HMM and FHMM

III. OUR SCHEME

Compared with traditional schemes, in this section, we proposed a differential privacy scheme which can get a better trade-off between utility and privacy. The key idea is to solve the problem of huge sensitivity through a specific transformation, achieve a novel result with a smaller noise and stronger privacy-preserving level.

*A. Notations*

In Table.1, the notations used in this paper are listed.

Table.1 Notations in our scheme

| Acronym | Descriptions |
|---|---|
| $\varepsilon$ | Privacy budget |
| $<D, D'>$ | Adjacent datasets |
| $\pi_i$ | Initial probability of appliance i |
| $A_i$ | Transition probability of appliance i |
| $B_i$ | Emission probability of appliance i |
| $N$ | The number of appliances |
| $X_i$ | Hidden states sequence of appliance i |
| $Y_i$ | Observed states sequence of appliance i |
| $Y_{sum}$ | Aggregate observed state sequence |
| $Y_{train}$ | Training energy consumption data |
| $x_t^{(i)}$ | Hidden state of appliance i at time t |
| $x_t^{(i)}{}'$ | Obfuscated hidden state of appliance i at time t |
| $y_t^{(i)}$ | Observed state of appliance i at time t |
| $y_t^{(i)}{}'$ | Obfuscated observed state of appliance i at time t |
| $y_t'$ | The aggregation of the obfuscated observed state at time t |
| $\lambda$ | The set of parameters $A, B, \pi$ |
| $GS_f$ | Global sensitivity |
| $LS_f$ | Local sensitivity |

*B. Design goal*

Traditional differential privacy proposed the notion for general statistical data sets. But for smart grid, the object of the protection is not only statistical data. Before Barbosa, there is no well-accepted rigorous definition of privacy in the smart grid environment. Barbosa [24] described it as that a consumption profile is a set of appliances, we say profiles P1 and P2 differ in at most one element if one is a proper subset of the other and the larger dataset profile contains just one additional appliance.

Base on this definition, we have made fine-grained improvements. We say profiles $D$ and $D'$ differing in at most one element if one is a proper subset of the other and the larger dataset profile contains just one state of an appliance.

Instantiated by the notion of differential privacy proposed by Dwork [27] and Barbosa [24], we propose the notion of differential privacy for datasets of the behavior signatures. Therefore, the adversary learns the similar information when there is a difference of the behavior signatures.

We call switch D and $D'$ differing in at most one element adjacent datasets, if the differential element is an additional behavior signature.

**Definition.1. Adjacent datasets**

M is a randomized algorithm. M satisfies $\varepsilon$-differential privacy if both of the datasets satisfy the following condition:

$$P_r\{M(D_i) \in S\} \leq e^\varepsilon \times P_r\{M(D') \in S\} \quad (1)$$

For all profiles $D$ and $D'$ differing in at most one state of an appliance.

**Definition.2. Global sensitivity**

For a mapping $f: D \rightarrow R^k$, $R^k$ denotes a k-dimensional vector. $D$ and $D'$ are an arbitrary pair of adjacent datasets. The global sensitivity of $f$ is

$$GS_f = \max_{D,D'} \| f(D) - f(D') \|_1 \quad (2)$$

For all the D and $D'$ differing in one appliance's switch state.

The design goals of the proposed scheme are given as follows. Inherited from Barbosa's [24] design goals, our schemes focus on the following aspects:

1) Enabling the calculation of the total consumption of a consumer over a period of time (*e.g.*, monthly billing);

2) Enabling the calculation of the total consumption of all consumers in a region at a certain instant of time;

3) Avoiding the measurement of the instantaneous consumption of an individual consumer at a certain instant of time.

Besides, we also propose two new design goals.

1) The entropy of the final obfuscated data should not be far from the original data.

2) There is no outlier in the final obfuscated load profile.

Figure.4 is quoted from Acs's scheme [30]. As shown in the obfuscated load profile, there are several values extremely lower than zero, which is against the common energy consumption behavior. We take these values as outliers.

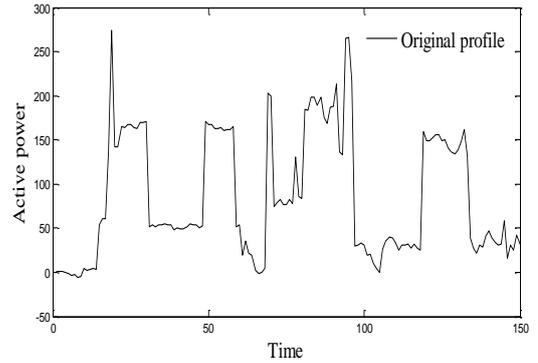

(a) Original load profile

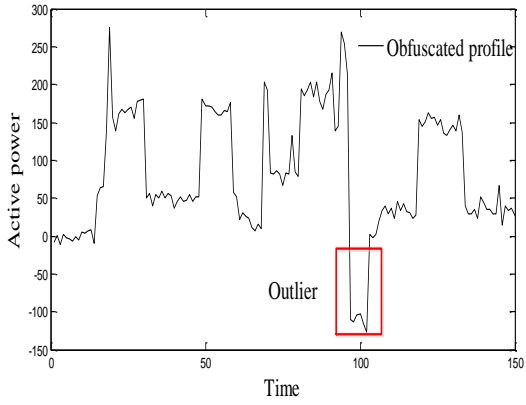

(b) Obfuscated load profile

Figure.4 Load profiles of original and obfuscated data

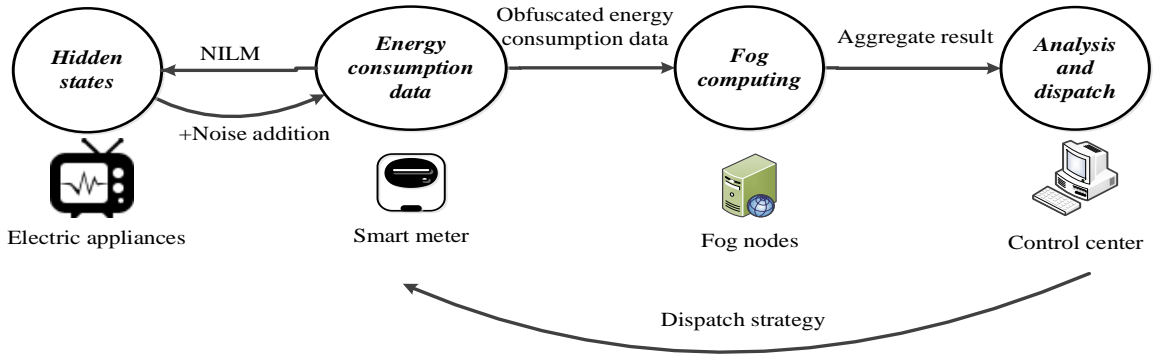

Figure.5 The system model of our scheme

## C. System model

We show the system model of our scheme in Figure.5 and show the architecture in Figure.6. The load signature is extracted from the energy consumption data. Then, each appliance's switch state related with the consumer's behavior is estimated based on the FHMM. Differing from the traditional differential privacy schemes which add noise into active power data, we add noise into the consumer's behaviors (the switch states of appliances) to implement the differential privacy. Then, the obfuscated energy consumption data is sent to the fog by smart meters. After processing the data with the fog computing nodes in groups, the aggregators will send the processed result to the cloud for further analysis.

## D. Appliance Modelling

As we have analyzed before, the energy consumption behavior can be modeled by the FHMM, in which the aggregated active power sequence of the entire appliances is regarded as the observed state, and the switch state sequence of each appliance is regarded as the hidden state. To estimate the hidden state, we need to estimate the related parameters in FHMM first.

We use $H$ to denote the set of switch states $H = \{OFF, ON_1, ... ON_\omega\}$. $ON_i$ represents the kind of switch state in ON-state. The related parameters of appliance $i$ in FHMM contain the initial probabilities $\pi_i = P(x^{(i)}_1)$, the conditional probabilities $A_i = P(x^{(i)}_t | x^{(i)}_{t-1})$, and the emission probabilities $B_i = P(y^{(i)}_t | x^{(i)}_t)$. To simplify the analysis, we use $\lambda_i$ to denote the set of parameters.

Based on the related parameters of appliance $i$, we can calculate the initial probability $\pi_i$, the transition probability $A_i$, the emission probability $B_i$ and the conditional probability of switch state $P(Y, X | \lambda_i, 1 \le i \le N)$ as follows:

$$\pi = \prod_{i=1}^{N} \pi_i = \prod_{i=1}^{N} P(x^{(i)}_1)$$

$$A = \prod_{i=1}^{N} A_i = \prod_{i=1}^{N} P(x^{(i)}_t | x^{(i)}_{t-1})$$

$$B = \prod_{i=1}^{N} B_i = \prod_{i=1}^{N} P(y^{(i)}_t | x^{(i)}_t)$$

$$P(Y, X | \lambda_i, 1 \le i \le N) = \pi A B$$

(3)

$N$ denotes the number of appliances. Expectation Maximization algorithm (*EM*) is a common solution to estimate these parameters by using an auxiliary function until the convergence to a local maximum occurs. In our paper, we don't adopt *EM*, instead, we take partial energy consumption data from all kinds of appliances as the training data $Y_{train}$ and estimate the parameters by Maximum Likelihood Estimation. The process is shown in Table.2.

Given a series of energy consumption data $Y_{sum}$ from a smart meter, we can estimate all the appliances' switch state sequences based on our FHMM model. With the Maximum Likelihood Estimation, we can estimate all the appliances' switch state sequences as follows:

$$X_1 X_2 ... X_N = \arg\max P(Y_{sum}, X | \lambda) \quad (4)$$

Here, argmax() represents the Maximum Likelihood Estimation of related parameters in our FHMM model.

As the hidden states of each appliance are easier to disclose user's habits and behaviors, adding noise into the hidden states is more effective to preserve user's privacy. Besides, noisy hidden state has better performance than the noisy energy consumption data in terms of data-utility. We show the detailed process in the next section.

**Algorithm 1. Estimate the switch state by FHMM**

**Input:** $Y_{train}$, $Y_{sum}$

**Output:** The switch state sequences of each appliance

(1) Input $Y_{train}$ into the FHMM as the training data.

(2) Calculate the Maximum Likelihood Estimation of $\lambda$

(3) Input $Y_{sum}$

(4) Calculate the switch state sequences based on $\lambda$

(5) Output switch states sequences

Table.2 Estimate the switch state by FHMM

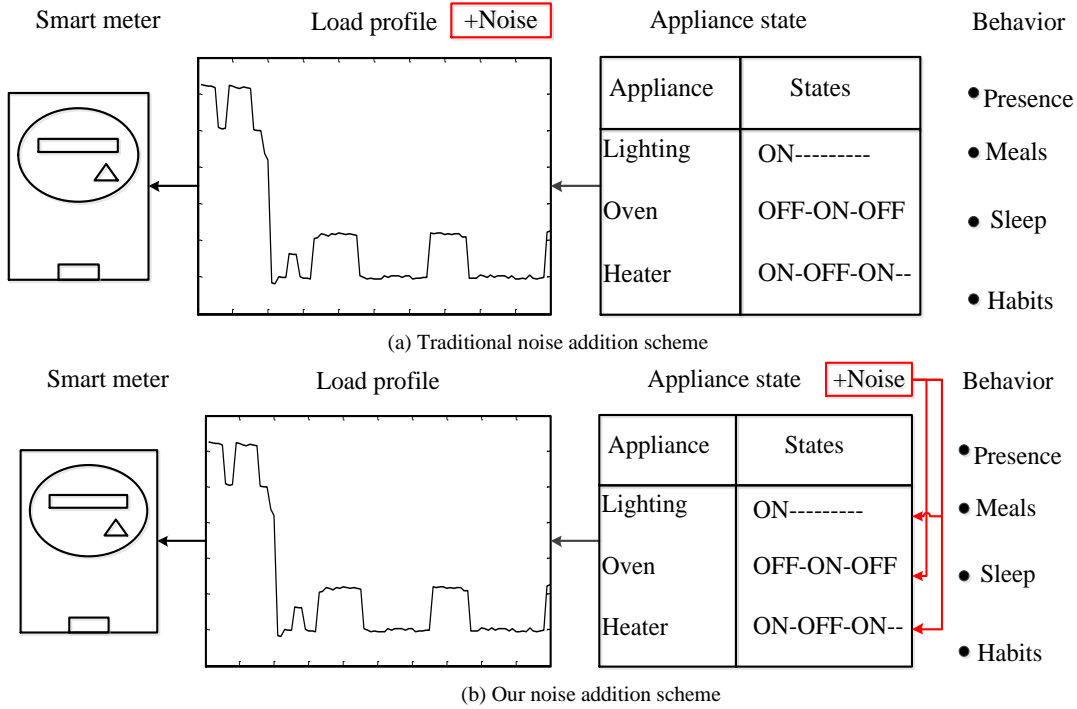

(a) Traditional noise addition scheme

(b) Our noise addition scheme

Figure.6 The difference of noise addition between traditional scheme and ours

*E. Noise addition*

**Definition.3. Local sensitivity**

For a mapping $f: D \to R^k$, in which $R^k$ denotes a k-dimensional vector and D' is an arbitrary adjacent dataset of D, the local sensitivity of f is

$$LS_f = \max_D \| f(D) - f(D') \|_1 \quad (5)$$

**Definition.4. Smooth sensitivity**

For a mapping $f: D \to R^k$, in which $R^k$ denotes a k-dimensional vector and D' is an arbitrary adjacent dataset of D, the local sensitivity of f is

$$S(D) = \max(Ls(D') \times e^{-\beta |D \Delta D'|}) \quad (6)$$

**Theorem.1.** For $f: D \to R^k$, the mechanism that adds noise with distribution $Lap(\Delta f / \varepsilon)$ provides $\varepsilon$-differential privacy.

The theorem has been proved by Dwork [28]. In this paper, $R^k$ represents the active power of all the appliances in a period of time. $f$ represents the process of FHMM. $\Delta f = S(D)$ and represents the maximum difference of the appliance's switch states for the two adjacent datasets.

After getting the switch state sequences of appliance $i$ in time $t$, we add Laplace into the switch states of each appliance to generate the obfuscated switch state $x^{(i)}_t{}'$. The detailed process can be expressed as follows:

$$x^{(i)}_t{}' = x^{(i)}_t + lap(S(D)/\varepsilon) \quad (7)$$

## F. Data re-aggregation

After we get the obfuscated switch state sequence, we can generate the obfuscated active power sequence based on the FHMM. While, considering the data-utility, we adjust the obfuscated active power as follows:

$$x^{(i)}_t{}' = \begin{cases} OFF & x^{(i)}_t{}' = 0 \\ ON_1 & x^{(i)}_t{}' = 1 \\ ... & ... \\ ON_w & x^{(i)}_t{}' = \omega \end{cases} \quad (8)$$

$$y^{(i)}_t{}' = \begin{cases} y^{(i)}_t & x^{(i)}_t{}' = x^{(i)}_t \cap x^{(i)}_t{}' \neq 0 \\ 0 & x^{(i)}_t{}' = 0 \\ CP & x^{(i)}_t{}' \neq x^{(i)}_t \cap x^{(i)}_t{}' \neq 0 \end{cases} \quad (9)$$

When $x^{(i)}_t{}' = x^{(i)}_t$ and $x^{(i)}_t{}' \neq 0$, the obfuscated active power based on FHMM is similar to the average value of the energy consumption data in total time slots. To reflect the real energy consumption, we take the original energy consumption data as the obfuscated active power in this time slot.

When $x^{(i)}_t{}' = 0$, theoretically, the obfuscated active power should be zero. However, as the relationship between the switch states and observed states is estimated by FHMM and may be nonzero when $x^{(i)}_t{}' = 0$. Therefore, we set $y^{(i)}_t{}' = 0$ in this situation.

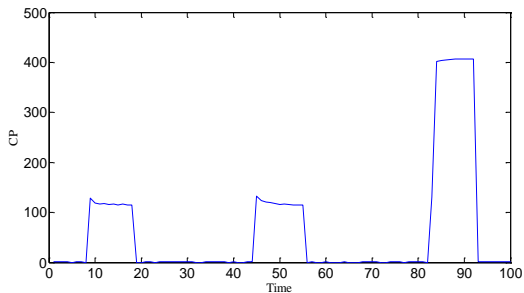

(a) Fridge

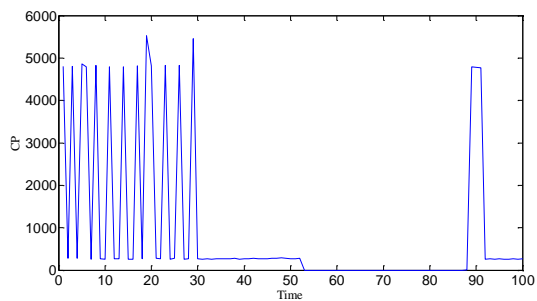

(b) Washer dryer

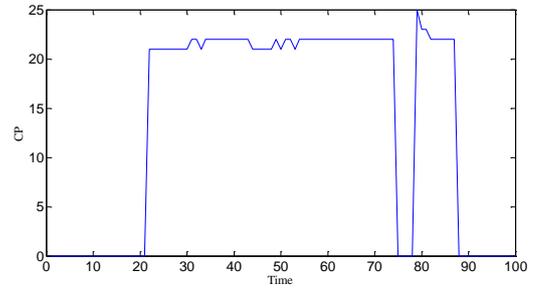

(c) Light

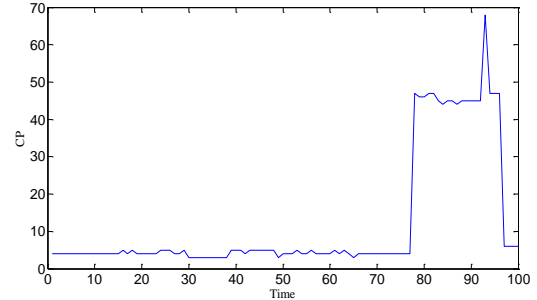

(d) Microwave

Figure.7 Energy consumption profile of each appliance

In fact, the active power of an appliance is a little different even in the same switch state. When $x^{(i)}_t{}' \neq x^{(i)}_t$ and $x^{(i)}_t{}' \neq 0$, to reflect the real energy consumption, we take the value from the Consumption Profile (*CP*) whose switch state is equal to $x^{(i)}_t{}'$ as our obfuscated active power. Figure.7 shows the energy consumption profiles of a fridge, a washer dryer, a light and a microwave, respectively. All these appliances have two or multiple states and the energy consumption dynamics are different even in the same operation mode. The final aggregated active power in time *t* can be calculated as follows:

$$y_t' = \sum_{i=1}^{N} y^{(i)}_t{}' \quad (10)$$

## IV. SECURITY ANALYSIS

### A. Privacy analysis

**Theorem 2: Our scheme provides $\varepsilon$-differential privacy.**

**Proof:**

The process of our scheme $T_{sum}$ can be expressed as follows:

$$T : Y \xrightarrow{FHMM} x^{(i)}$$

As the mapping $T_b$ representing the FHMM can be regarded as a linier mapping approximately and $|E \oplus F|$ represent the number of different elements between *E* and *F*.

We use $T_{sum}$ to denote the process of FHMM which could be regarded as a combination of two sub-processes $T_a$ and $T_b$. $T_a$ represents the transformation from each appliance's obfuscated switch states to the power consumption data. $T_b$ represents the processes of data disaggregation.

Therefore, the process of FHMM can be seen as a combination $T_{sum} = T_b(T_a(X_1), T_a(X_2)...T_a(X_n))$.

1) According to the analysis by Dwork [26], adding Laplace noise into the original switch states provides $\varepsilon$-differential privacy. We use M to denote this process.

2) As the map between appliances' obfuscated switch states and the power consumption data is one to one.

We use T to denote this process.

$T(E) = \pi_E A_E B_E$

$T(F) = \pi_F A_F B_F$

If $a \in A, b \in B$ and $a - b \vert < e$, we have $a \oplus b \approx 0$.

if $E \oplus F = 0$,

$\pi_E \oplus \pi_F = 0$

$A_E \oplus A_F = 0$

$B_E \oplus B_F = 0$

Therefore $T(E) \oplus T(F) = 0$

And $|T(E) \oplus T(F)| \le 1 \times |E \oplus F|$.

Thus, the map $T$ is c-stable and the value of c is one.

3) According to the property of c-stable, the transformation $T_a$ which is equivalent to $M \circ T$ provides $\varepsilon$-differential privacy.

4) According to the combinability property of differential privacy, our scheme provides $\varepsilon$-differential privacy.

*B. Utility analysis*

*Definition.5.* $(\alpha, \beta)$ **-utility:**

For two datasets $D, D'$ which represent the original data and processed data, a randomized function Q satisfies $(\alpha, \beta)$-utility if it has the following property

$$\Pr[\|Q(D')-Q(D)\|_1 \le \alpha] > 1 - \beta \quad (11)$$

Here, $\alpha$ represents the upper bound of distance between $Q(D')$ and $Q(D)$. The smaller $\alpha$ is, the higher the level of data-utility will be achieved. $\beta$ measures the probability $\Pr[\|Q(D')-Q(D)\|_1 \le \alpha]$, and the probability of the distance between $Q(D')$ and $Q(D)$ below the upper bound increases when the value of $\beta$ decreases.

**Theorem 3: our scheme satisfies** $(\alpha, \beta)$ **-utility.**

**Proof:**

In this paper, D can be regarded as the real switch states of an appliance in different time slots. $D'$ can be regarded as the noisy switch states. Q represents the FHMM algorithm, which decides the mapping between the observed state and switch state.

1) As the mapping between the observed state and switch state in the FHMM can be seen as a linear mapping, we have

$$Q(D') = kD' + b \quad (12)$$

$$Q(D) = kD + b \quad (13)$$

$$D' = D + \Delta D \quad (14)$$

$k$ and $b$ are linear parameters and $\delta$ represent the correction coefficient.

2) Based on the above analysis and the property of norm, we have

$$\begin{aligned}\|Q(D')-Q(D)\|_1 &= k \| (D' - D) \|_1 \\ &= k \| \Delta D \|_1 = k \sum_i^n \Delta d_i\end{aligned} \quad (15)$$

3) As each noise $\Delta d_i \sim laplace(0, \frac{\Delta f}{\varepsilon})$, the value of $\sum_i^n \Delta d_i$ will converge to zero when the value of n is large enough. Therefore, we have

$$\sum_i^n \Delta d_i < \theta \quad (16)$$

$$\|Q(D')-Q(D)\|_1 \le k\theta \quad (17)$$

4) According to the above analysis, we have

$$\Pr[\|Q(D')-Q(D)\|_1 \le k\theta] \ge 1 \quad (18)$$

Therefore, our scheme is proved to satisfy $(k\theta, 0)$-utility.

V. PERFORMANCE EVALUATION

In this section, we use F1-score [38] to measure the performance of our scheme in terms of the level of privacy-preservation and adopt Kullback–Leibler divergence [45] to measure the level of data-utility based on the REDD data set [40] with the tool NILMTK [41]. Then, we compare our scheme with Barbosa's scheme and Sankar's scheme as follows.

*A. Privacy-preserving level of our scheme*

As we know, F1-score is an efficient metric to measure the level of privacy preservation, which can be regarded as a broadly accepted measuring tool for the accuracy of NILM F1-score is widely used in NILM and completed in NILMTK [41].

F1-score can be calculated as follows:

$$F1\text{-score} = \frac{2 \times Precision \times Recall}{Precision + Recall} \quad (19)$$

Here, *Precision* and *Recall* represent the positive predictive value and the recall sensitivity respectively. They can be calculated as follows:

$$precision = \frac{T_P}{T_P + F_P} \times 100\% \quad (20)$$

$$recall = \frac{T_P}{T_P + F_N} \times 100\% \qquad (21)$$

$T_P$ represents the value of true positive which means the number of appliances that are correctly predicted to be on. $F_P$ represents the value of false positive which means the number of appliances that are wrongly predicted to be on. $F_N$ represents the value of false negative which means the number of appliances that are wrongly predicted to be off. When the F1-score goes high, the application usage patterns can be tracked more accurately.

We adopt FHMM to estimate the switch states from the active power data obfuscated by Barbosa's scheme, Sankar's scheme and our scheme. The F1-scores of different schemes are shown in Figure.8.

After the noise addition, the F1-score by NILM has fallen. We can find that the F1-score of our scheme is smaller than the other schemes, which means that our scheme has a stronger advantage resisting NILM. When the same noise is adding to the electricity consumption data, the F1-score of our scheme by NILM decreases greatly. So our scheme has a higher level of privacy preservation.

### B. Data-utility of our scheme

Kullback–Leibler divergence is a measure of how one probability distribution diverges from another expected probability distribution. The definition is shown as follows

$$D(P \| Q) = \sum_i P(i) \log \frac{P(i)}{Q(i)} \qquad (22)$$

$P$ represents the original discrete probability distribution and $Q$ represents the fitting distribution. The larger the Kullback–Leibler divergence is, the larger the difference of the two distributions is. In this section, the original energy consumption data serves as $P$ and the obfuscation data serves as $Q$. When the Kullback–Leibler divergence grows higher, the difference of original data and obfuscation data will grow higher as well, and the level of data-utility will become lower.

At last, the multiple appliances' energy consumption data processed by different schemes are shown in Figure.9. We could find directly that our scheme has a lower impact of data utility.

We show the profiles of Kullback–Leibler divergence based on different $\varepsilon$ values in Figure.10. Through comparing those figures, we can find that the energy consumption profile processed by our scheme is very close to the original energy consumption profile. Besides, by observing the curves of Kullback–Leibler divergence in Figure.10, we can find that the Kullback–Leibler divergence decreases with the value of $\varepsilon$ and our scheme's Kullback–Leibler divergence is smaller than the Barbosa's scheme and Sankar's scheme. We take the value of $\varepsilon$ from the discrete set $\{0.1,1,5,10\}$ and calculate the Kullback–Leibler divergences of different schemes based on different $\varepsilon$. According to the property of Kullback–Leibler divergence, the Kullback–Leibler divergence will decrease when $\varepsilon$ decreases and the scheme with a lower Kullback–Leibler divergence provides a higher level of data-utility. Therefore, our scheme has an obvious advantage in data-utility.

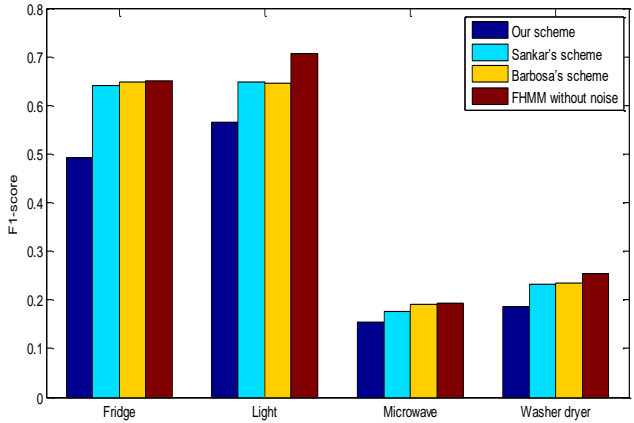

Figure.8. The F1-scores of different schemes

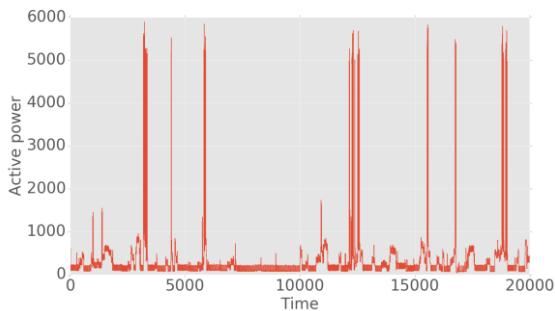

(a) Original data of multiple applications

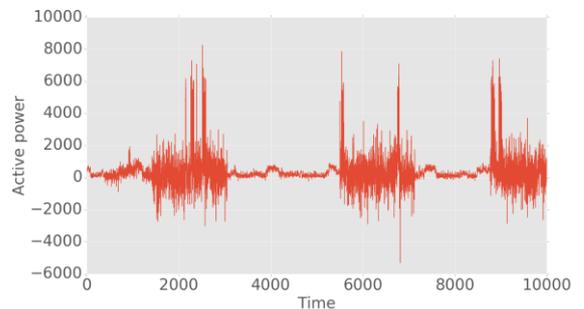

(b) Data obfuscated by Barbosa

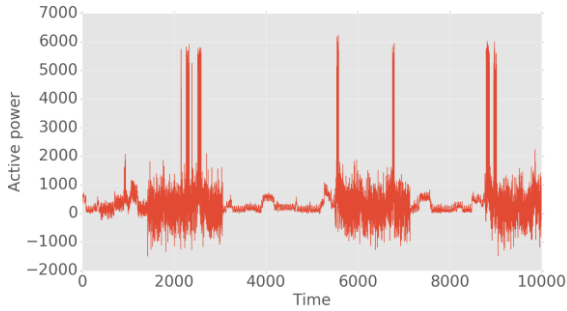
(c)Data obfuscated by Sankar

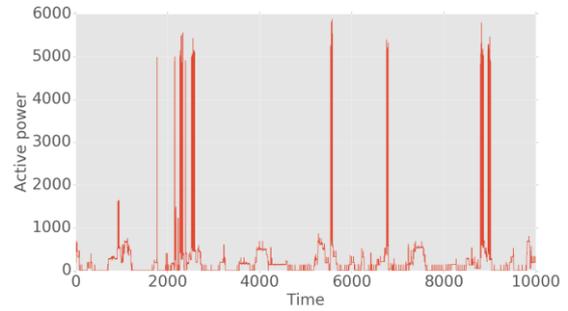
(d)Data obfuscated by our scheme

Figure.9 Energy consumption profiles processed by different schemes (multiple appliances ε = 5)

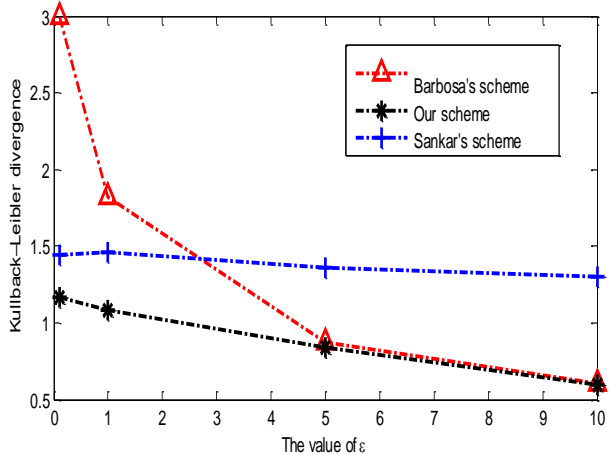
Figure.10. Kullback–Leibler divergence of different schemes

## VI. CONCLUSION

In this paper, we propose a privacy-preserving scheme based on the obfuscated switch states to realize the differential privacy towards fog computing in smart grid. We adopt the Factorial Hidden Markov Model to estimate the switch states of each appliance. Then, noise following Laplace distribution is added into the switch state to achieve the differential privacy. Based on the obfuscated switch states, we generate the obfuscated observed states and adjust them to guarantee the data-utility. Therefore, the appliance energy consumption patterns can be masked, even if the adversary can obtain the near real-time load profile. At last, we analyze the performance of our scheme, and compare it with other similar schemes in terms of the level of privacy-preserving (F1-score) and data-utility (Kullback–Leibler divergence). The security analysis and performance evaluation show that our scheme provides a better utility-privacy tradeoff. In the future, we will focus on extending the FHMM to further preserve user's privacy without compromising the data-utility. In addition, the limitation of the FHMM algorithm is its high computational cost. We will try to design the lightweight FHMM based method that is suitable for fog clients.

## ACKNOWLEDGMENT

This work is partially supported by Natural Science Foundation of China under grant 61402171, and the Fundamental Research Funds for the Central Universities under grant 2016MS29.